\documentclass[aps,prl,superscriptaddress,twocolumn]{revtex4-2}
\usepackage[pdftex]{graphicx}
\usepackage{amsfonts}
\usepackage{amsthm}
\usepackage{amsmath}
\usepackage{amsfonts}
\usepackage{amssymb}
\usepackage{bm}
\usepackage{braket}
\usepackage[normalem]{ulem}
\usepackage[colorlinks=True,linkcolor=red,citecolor=blue,urlcolor=blue]{hyperref}
\usepackage[dvipsname]{xcolor}
\newcommand\eq[1]{\begin{align}#1\end{align}}

\newcommand{\deltat}{\Delta_\mathrm{typ}}
\newcommand{\yt}{y_\mathrm{typ}}
\newcommand{\w}{\omega}
\newcommand{\pd}{\phantom\dagger}
\newcommand{\comment}[1]{}

\newcommand{\bel}[2][\mathrm{l}]{\omega^{\scriptscriptstyle(#1)}_{#2}}
\newcommand{\bed}[2][\mathrm{e}]{\omega^{\scriptscriptstyle(#1)}_{#2}}
\newcommand{\beld}[2][\mathrm{l/e}]{\omega^{\scriptscriptstyle(#1)}_{#2}}

\begin{document}
\title{Self-consistent theory of mobility edges in quasiperiodic chains}

\author{Alexander Duthie}
\email{alexander.duthie@chem.ox.ac.uk}
\affiliation{Physical and Theoretical Chemistry, Oxford University,
South Parks Road, Oxford OX1 3QZ, United Kingdom}

\author{Sthitadhi Roy}
\email{sthitadhi.roy@chem.ox.ac.uk}
\affiliation{Physical and Theoretical Chemistry, Oxford University,
South Parks Road, Oxford OX1 3QZ, United Kingdom}
\affiliation{Rudolf Peierls Centre for Theoretical Physics, Clarendon Laboratory, Oxford University, Parks Road, Oxford OX1 3PU, United Kingdom}

\author{David E. Logan}
\email{david.logan@chem.ox.ac.uk}
\affiliation{Physical and Theoretical Chemistry, Oxford University,
South Parks Road, Oxford OX1 3QZ, United Kingdom}
\affiliation{Department of Physics, Indian Institute of Science, Bangalore 560012, India}

\date{\today}
\begin{abstract}
We introduce a self-consistent theory of mobility edges in nearest-neighbour tight-binding chains with quasiperiodic potentials. Demarcating boundaries between localised and extended states in the space of system parameters and energy, mobility edges are generic in quasiperiodic systems which lack the energy-independent self-duality of the commonly studied Aubry-Andr\'e-Harper model. The potentials in such systems are strongly and infinite-range correlated, reflecting their deterministic nature and rendering the problem distinct from that of disordered systems. Importantly, the underlying theoretical framework introduced is model-independent,  thus allowing analytical extraction of mobility edge trajectories for arbitrary quasiperiodic systems. We exemplify the theory using two families of models, and show the results to be in very good agreement with the exactly known mobility edges as well numerical results obtained from exact diagonalisation.
\end{abstract}

\maketitle

The phenomenon of Anderson localisation~\cite{anderson1958absence} is conventionally discussed in the context of disordered quantum systems. Quenched randomness is not however a prerequisite for localisation. Indeed there exists a family of systems --
those with \emph{quasiperiodicity} -- which are non-random and deterministic, yet host localisation. 
The simplest and arguably most famous member of the family, the Aubry-Andr\'e-Harper (AAH) model~\cite{aubry1980analyticity,Harper_1955}, hosts a localisation transition~\cite{aubry1980analyticity} already in one dimension. Quasiperiodic 
chains also commonly show other interesting phenomena such as mobility edges, multifractal eigenstates both at and away from criticality, and ``mixed phases'' with both extended and localised eigenstates~\cite{prange1983wave,sarma1988mobility,sarma1990localization,biddle2009localization,biddle2010predicted,biddle2011localization,ganeshan2015nearest,wang2020onedimensional,boers2007mobility,li2017mobility,gopalakrishnan2017self,roy2018multifractality}; and are readily implemented in experimental quantum emulators with ultracold atoms~\cite{luschen2018single,an2020observation}.

While elusive in $d\leq 2$-dimensional, short-ranged disordered systems, mobility edges (ME) which demarcate the boundary between localised and extended states in the space of Hamiltonian parameters and energy, are commonplace and essentially generic in quasiperiodic systems. The aforementioned AAH model is in effect a special  case, where all eigenstates undergo a localisation transition at a unique critical value of the quasiperiodic potential strength, and MEs cease to exist~\cite{aubry1980analyticity}. This is due to an exact duality in the model which is independent of energy. Any distortion of the model which breaks this duality typically leads to a genuine mobility edge in the spectrum.

Previous successes in understanding mobility edges in such systems have tended to focus on ideas specific to particular models, such as energy-dependent generalised duality transformations~\cite{biddle2010predicted,biddle2011localization,ganeshan2015nearest} or continuum models with bichromatic incommensurate potentials~\cite{boers2007mobility,li2017mobility}.
The propensity of quasiperiodic models to possess MEs naturally means a theoretical framework to predict and understand them, which is model-independent, is of  basic importance; and this constitutes the central motivation of the present work.

We introduce a self-consistent theory of mobility edges in quasiperiodic systems based on the analysis of the local propagator, $G_j(t)=-i\Theta(t)\braket{j|e^{-iHt}|j}$, which physically measures the return probability amplitude of a state initialised at site $j$. The propagator is analysed in the energy ($\omega$) domain, wherein it acquires a self-energy whose imaginary part, $\Delta_j(\omega)$, is the central quantity of interest.  Physically, $\Delta_j(\omega)$ is proportional to
the rate of loss of probability from site $j$ into states of energy $\omega$, and is thus a natural diagnostic for localisation or its absence. The characteristics of $\Delta_j(\omega)$ have in fact long proven successful in understanding Anderson transitions~\cite{anderson1958absence,abou-chacra1973self,economous1972existence,ThoulessReview1974,licciardello1975study,logan1985anderson,logan1987dephasing}.  However, much of the analytical progress there was rendered possible by the independence of the random site-energies, and consequent independence of the local self-energies. Quasiperiodic systems in this regard present a significant challenge, as the deterministic nature of the potential means the site-energies and self-energies are strongly and infinite-range correlated.

As concrete models for establishing and testing the theory, we consider one-dimensional nearest-neighbour tight-binding Hamiltonians of form
\eq{
	H=V\sum_{j=0}^{L-1}\epsilon_j^{\phantom\dagger} c_j^\dagger c_j^{\phantom\dagger} +
	J\sum_{j=0}^{L-2}[c_j^\dagger c_{j+1}^{\phantom\dagger}+ \mathrm{H.c.}] \,, 
	\label{eq:ham}
}
where $\epsilon_j$ encodes the quasiperiodic potential specific to the model 
(for specificity we consider $V,J\geq 0$, and unless stated otherwise take $J=1$).
In particular, we consider two families of models. The first, referred to as the $\beta$-models, is described by~\cite{ganeshan2015nearest}
\eq{
	\epsilon_j^{\pd} = \cos(2\pi\kappa j +\phi)[1-\beta\cos(2\pi\kappa j +\phi)]^{-1}\,,
	\label{eq:beta-model}
}
with an irrational $\kappa$ (chosen as the golden mean) reflecting the quasiperiodicity, and $\phi \in [0,2\pi]$ a global phase shift used to accumulate statistics. The $\beta$-models are self-dual, and host a single ME given  by $\omega_{\mathrm{ME}} = (2-V)/\beta$~\cite{ganeshan2015nearest}.  Note that the standard AAH model is recovered as the $\beta =0$ limit, where the ME becomes a vertical line parallel to the $\omega$-axis at $V=2$, indicating that all states undergo a localisation transition at $V=2$.

The second family is the so-called \emph{mosaic}-models~\cite{wang2020onedimensional} parametrised by an integer $l$. These 
non-self-dual models have an AAH potential on every $l^\mathrm{th}$ site, while all remaining sites have $\epsilon_l=0$. Formally,
\eq{
	\epsilon_j^{\pd} = \begin{cases}
					\cos(2\pi\kappa j +\phi)~~~:~j=lk \\
					0~~~~~~~~~~~~~~~~~~~~:~\mathrm{otherwise}
				\end{cases}
	\label{eq:mosaic-model}
}
where $k\in\mathbb{Z}$. While MEs are known analytically for arbitrary $l$, for brevity we here consider explicitly the $l=2$ model, where
the spectrum hosts two MEs given by $\omega_\mathrm{ME}=\pm 2/V$~\cite{wang2020onedimensional}.

Our theory centres on analysis of the local self-energy, $S_j(\omega)$, defined via the local Green function on site $j$ as
\eq{
	G_j^{\pd}(\omega) = [\omega^+-V\epsilon_j-S_j(\omega)]^{-1}\,,
	\label{eq:Gdef}
}
where $\omega^+=\omega+i\eta$ with $\eta=0^+$, and $S_j=X_j-i\Delta_j$.  We focus on the imaginary part $\Delta_{j}=-\mathrm{Im}S_{j}$, since it serves as a probabilistic order parameter for a localisation-transition: $\Delta_j$ is [non-]vanishing with unit probability in 
an [extended] localised phase~\cite{abou-chacra1973self}. For a one-dimensional  nearest-neighbour model $S_j(\omega)$ can be expressed as
\eq{
	S_j^{\pd}(\omega) = J^2[G_{j-1}^{(j)}(\omega)+G_{j+1}^{(j)}(\omega)]\,,
	\label{eq:Sjdecomp}
}
with $G_{j\pm1}^{(j)}$ the local propagator for site $j\pm1$ with site $j$ removed. As the local self-energy is a sum of two end-site propagators of a semi-infinite chain, localisation or its absence can  be inferred from the properties of end-site propagators alone~\cite{abou-chacra1973self};  we thus focus in the following on the self-energy of an end site, denoted $S_0(\omega)$. Since $S_0 = J^2 G_1^{(0)}$, and the imaginary part of the Green function is proportional to the local density-of-states (LDoS), the typical value of $\Delta_0$ (denoted henceforth as $\deltat$) is proportional to the typical LDoS; indeed the latter was proposed as an order-parameter for localisation in both quasiperiodic~\cite{ganeshan2015nearest} and also disordered systems~\cite{logan1987dephasing,janssen1998statistics,dobrosavljevi2003typical}. This order parameter is finite in a extended phase and vanishes on approaching the transition from the extended side. In a localised phase by contrast, $\Delta_0 \propto\eta$ vanishes. This leads to a corresponding (and considerably less studied) order-parameter $\yt^{-1}=(\deltat/\eta)^{-1}$, which is finite in the localised phase and vanishes on approaching the transition from the localised side.
A mobility-edge is thus signalled by a \emph{simultaneous} vanishing of $\deltat$ from the extended side and divergence of $\yt$ from the localised side.

\begin{figure}[!t]
\includegraphics[width=\linewidth]{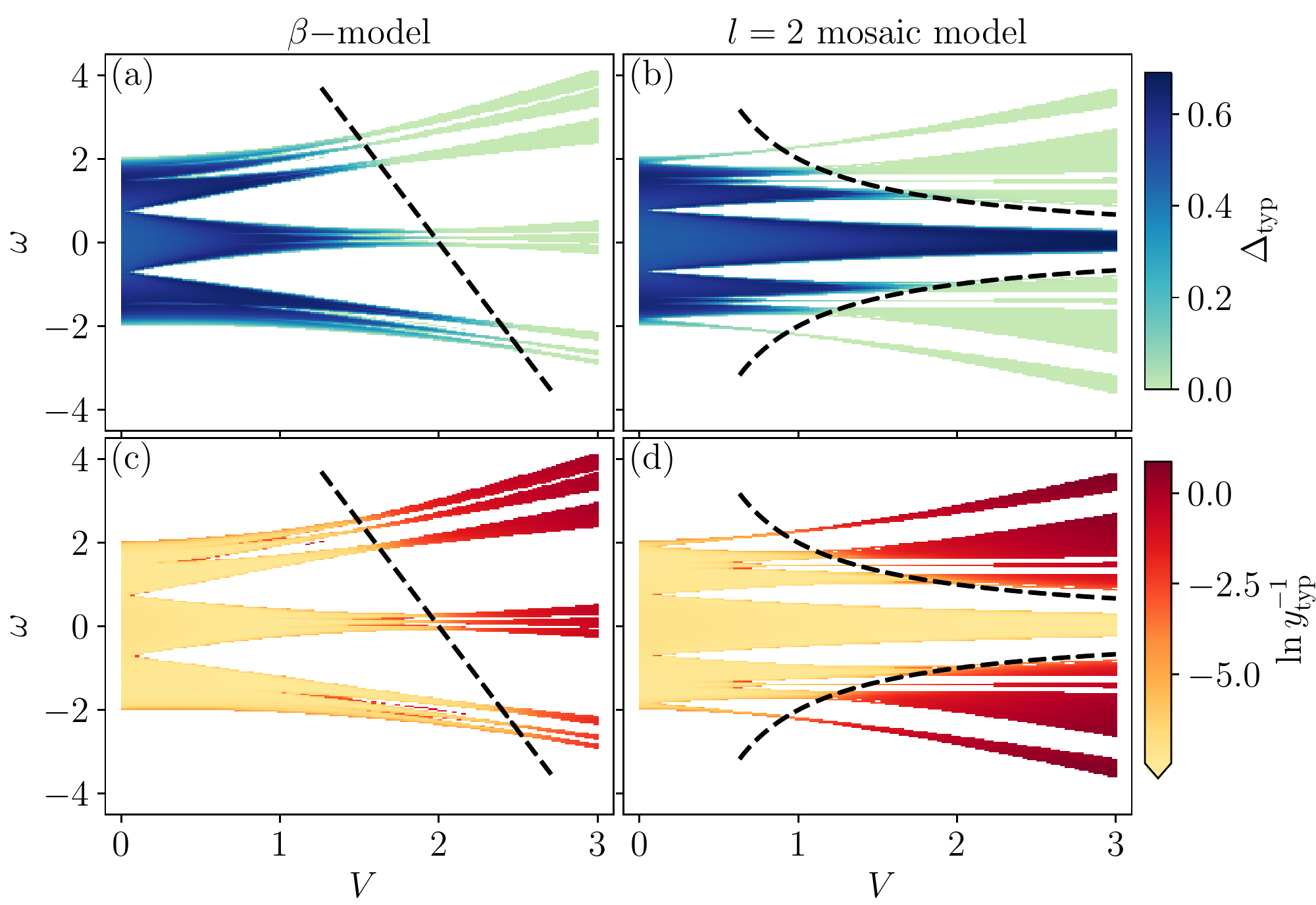}
\caption{
Spectrum of the $\beta$-model (for $\beta=0.2$) and the $l=2$ mosaic model, colour-coded with $\deltat(\omega)$ (panels 
(a), (b)) and $\ln\yt^{-1}(\omega)$ (panels (c), (d)). Black lines denote the known mobility edges~\cite{ganeshan2015nearest,wang2020onedimensional}. In panels (a), (b) a finite  $\deltat$ indicates the presence of extended states and a vanishingly small value indicates localised states. Concomitantly, in panels (c), (d), a finite $\yt^{-1}$ signals localised states whereas a vanishingly small $\yt^{-1}$  (divergent $\yt$ in the thermodynamic limit) signals extended states. These plots show  $\deltat$ to be a valid `order-parameter' for the localisation-transition/mobility-edge coming from the extended side, while $\yt^{-1}$ is its counterpart on approaching the transition from the localised side. Results obtained from exact diagonalisation with $L=2500$ sites and $\eta\propto 1/L$, with averaging over 5000 $\phi$-values. }
\label{fig:colour-plot-delta}
\end{figure}
 
In our theory we analyse both $\deltat$ and $\yt$. First, however, we show numerical results obtained from exact diagonalisation for the two models Eqs.~\eqref{eq:beta-model} and \eqref{eq:mosaic-model}. The local propagator of the end site can be computed as $G_0(\omega)=\braket{0|(\omega^+-H)^{-1}|0}$ with the matrix-inversion performed numerically (where $\ket{0}$ denotes a state localised on the end site), from which $\Delta_0$ can be computed using Eq.~\eqref{eq:Gdef}. $\deltat$ 
is then obtained as the geometric mean of the distribution of $\Delta_{0}$, by accumulating statistics over $\phi$.
Since $\eta$ should be on the order of the mean level spacing we take $\eta =c/L$ (with $c\sim \mathcal{O}(1)$~\footnote{Results are insensitive to the precise numerical prefactor.}), and $\yt$ is simply obtained as $\deltat/\eta$. The results are shown in Fig.~\ref{fig:colour-plot-delta}, where the spectrum of both models is plotted in the  $(V,\omega)$-plane with the data colour-coded according to $\deltat$ in panels (a)-(b) and $\ln\yt^{-1}$ in (c)-(d). The behaviour of the numerically obtained $\deltat$ and $\yt^{-1}$ -- in particular their drop to vanishing values on approaching the MEs from the extended and localised sides respectively -- clearly shows their utility as order-parameters. Having established that, we now turn to their theoretical analysis. 

We begin by using Eqs.~\eqref{eq:Gdef} and \eqref{eq:Sjdecomp} to express $S_0(\omega)= J^2 G_1^{(0)}$  as a continued fraction (CF),
\eq{
	S_0^{[n]}(\omega)=\cfrac{J^2}{\omega^+-V\epsilon_1-\cfrac{J^2}{\omega^+-V\epsilon_2-\cfrac{J^2}{\cdots-S_n^{[0]}(\omega)}}}\,,
	\label{eq:S0-contfrac}
}
where the superscript $[n]$ denotes that the CF has been continued (exactly) to order $n$~\footnote{Eq.~\eqref{eq:S0-contfrac} is formally the same as the CF running \textit{ad infinitum}, and thus remains exact.}. From this, $\Delta_0(\omega)$ and $y_0(\omega)$ can likewise be expressed as CFs.

At this stage, there are two conceptually distinct directions one can take. The first is to set the terminal self-energy in Eq.~\eqref{eq:S0-contfrac} to a typical value, $S_n^{[0]}\to-i\deltat$, and obtain a distribution of $\Delta_0^{[n]}$ over an ensemble of $\phi$-values. This distribution depends parametrically on $\deltat$.  Self-consistency is then imposed by requiring that $\deltat$ obtained from it is equal to the parametric $\deltat$.  This comprises a self-consistent mean-field theory at $n^\mathrm{th}$ order, in the spirit of the self-consistent theory of Anderson localisation~\cite{abou-chacra1973self}. The second,  along the lines of  Anderson's original work~\cite{anderson1958absence}, is to analyse the convergence of the CF for $y_{0}$. This converges with unit probability in the localised phase, such that $y_0$ is finite; while in the extended phase it fails to  converge, indicating a divergent $y_0$ in the thermodynamic limit. The convergence or lack thereof of the CF for $y_0$ is of course intimately connected to whether a finite or divergent self-consistent solution arises for $\yt$, which connects the two concepts~\cite{duthie2020localisation}.

In this work we take the first of these two directions, and analyse the theory explicitly at leading order, $n=1$ (dropping the superscript $[n]$ from now). From Eq.~\eqref{eq:S0-contfrac}, 
 \eq{
	\Delta_0^{\pd}(\omega) = \frac{J^2(\eta+\deltat)}{(\omega-V\epsilon_1)^2+(\eta+\deltat)^2}\,.
	\label{eq:delta0-leading}
}
In a localised regime, where $\deltat \propto \eta$, the relevant quantity $y_{0}=\Delta_{0}/\eta$ can be expressed as
\eq{
	y_0^{\pd}(\omega) = \frac{J^2(1+\yt)}{(\omega-V\epsilon_1)^2}\,.
	\label{eq:y0-leading}
}
With $\langle\langle\cdots\rangle\rangle$ denoting an average over $\phi$ and end sites~\footnote{The average over end sites is necessary to account for the different functional forms that end-site potentials may have, e.g.\ in the mosaic models~\cite{supp}.}, $\yt$ is self-consistently determined from $\ln \yt =\langle\langle \ln y_{0}\rangle\rangle$, and hence from Eq.\ \eqref{eq:y0-leading} by
\eq{
	\ln(1+\yt^{-1}) = \langle\langle\ln(\omega-V\epsilon_1)^2\rangle\rangle-\ln J^2\, .
	\label{eq:yt-leading}
}
In the extended regime, since $\deltat$ is finite, the limit $\eta\to0$ can be taken in Eq.~\eqref{eq:delta0-leading}, leading to
$\Delta_0 = J^2\deltat[(\omega-V\epsilon_1)^2+\deltat^2]^{-1}$;
which from $\ln\deltat =\langle\langle\ln \Delta_{0}\rangle\rangle$ yields the desired equation for the self-consistent $\deltat$,
\eq{
	\langle\langle \ln[(\omega-V\epsilon_1)^2+\deltat^2]\rangle\rangle-\ln J^2=0\,.
	\label{eq:deltat-leading}
}
Before proceeding, we lay out clearly how the MEs (along with the regions of localised and extended states) are diagnosed  using $\yt$ and $\deltat$.

At any point in the $(V,\omega)$-plane, a finite value of $\deltat(V,\omega)$ implies that states there are extended. If  by contrast $\yt(V,\omega)=\deltat/\eta$ is finite then states are localised, provided they exist for that $(V,\omega)$ point;  which requires $\omega$ to lie within the localised band edges $\bel{\pm}$ of the average DoS, itself given to leading-order by 
$\langle\langle \delta(\omega-V\epsilon_1)\rangle\rangle$~\cite{supp}. A point in the $(V,\omega)$-plane can thus be unambiguously identified as lying on a ME if (i) it lies in the interval $\omega \in [\bel{-},\bel{+}]$ and (ii) $\deltat$ and $\yt$ simultaneously  vanish and diverge (respectively) at that point. Additionally, the theory also predicts where localised and extended states may reside in the $(V,\omega)$-plane. Define $V_-$$(V_+)$ as the $V$ where the MEs enter(leave) the spectrum on increasing $V$ from 0, such that $V_{\mp}$ correspond  to the points where $\omega_{\mathrm{ME}} =\bel{\pm}$~\cite{supp}. Then for $V<V_-$ all states are extended, with spectral edges determined by the vanishing of $\deltat(\omega)$. For $V>V_+$ by contrast, all states are localised and the localised band edges form the spectral edges.

With this in hand, we turn to the results of our theory. Eqs.~\eqref{eq:yt-leading} and \eqref{eq:deltat-leading} are valid respectively on the localised and extended sides of a ME. Approaching the ME from the two sides amounts to $\yt^{-1}$ and $\deltat$ vanishing in Eqs.~\eqref{eq:yt-leading} and \eqref{eq:deltat-leading}. Reassuringly, both these conditions yield identical expressions for the self-consistent ME,
\eq{
	\langle\langle \ln[(\omega_\mathrm{ME}-V\epsilon_1)^2]\rangle\rangle-\ln J^2=0\,.
	\label{eq:ME-general}
}
Significantly, this is completely independent of the specific model. Such a model-independent theory of MEs is a central result of this work. Moreover, this also shows that it is sufficient to analyse $\yt \equiv \yt(V,\omega)$ to obtain the MEs.

\begin{figure}
\includegraphics[width=\linewidth]{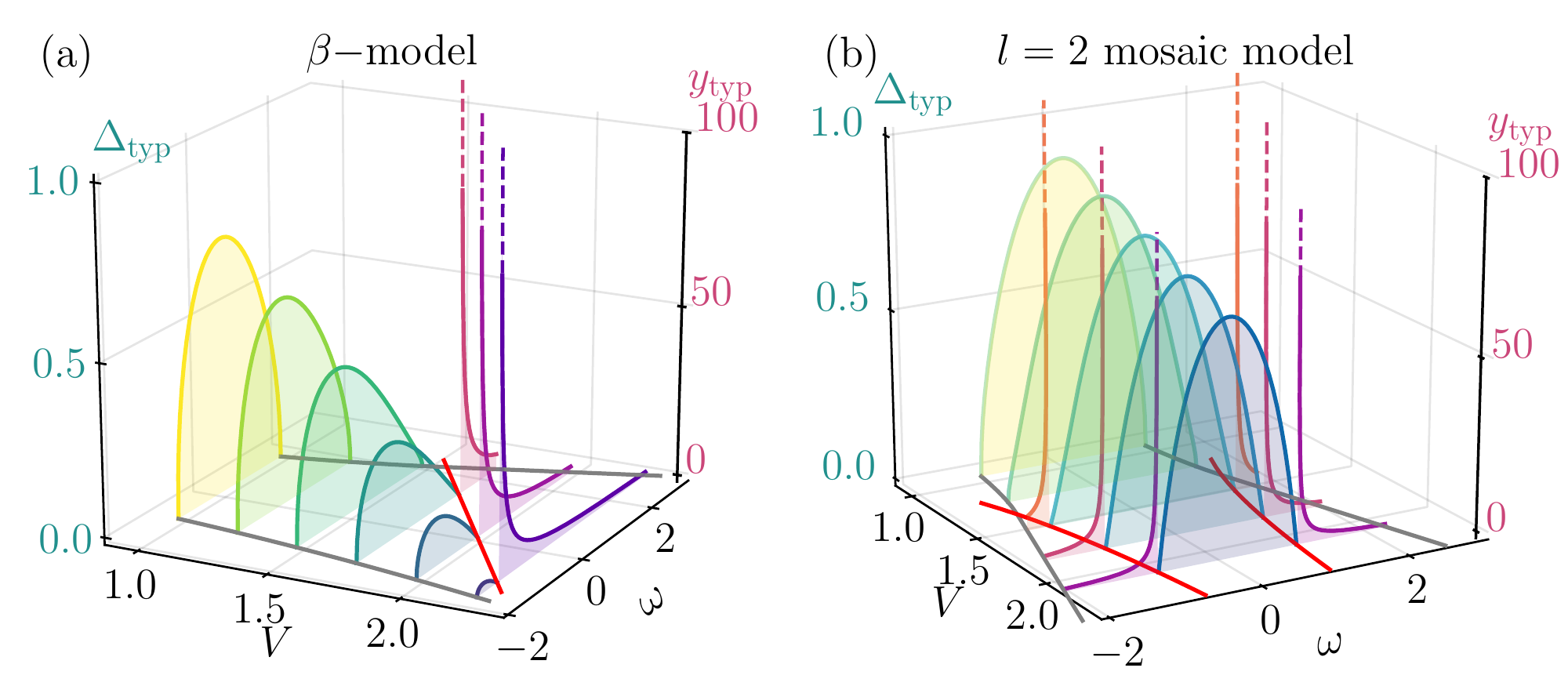}
\caption{Results for $\Delta_\mathrm{typ}$ and $y_\mathrm{typ}$ for (a) the $\beta$-model (with $\beta=0.2$) and (b) the 
$l=2$ mosaic model, obtained from the leading-order theory. A finite $\deltat$ implies the presence of extended states, while its vanishing indicates that states if present are localised. Within the edges of the spectrum  (shown as grey curves in the $(V,\omega)$ plane), the presence of localised states is indicated by a finite $\yt$. A simultaneous vanishing of $\deltat(\w)$ and divergence of $\yt(\w)$ signals the presence of mobility edges, which are denoted by the red curves in the $(V,\omega)$-plane. Note that the scales for $\deltat$ and $\yt$ in the plots are naturally very different.
}
\label{fig:Delta-y-typical-leading}
\end{figure}

Turning to specific results, we first discuss the $\beta$-model. Eq.~\eqref{eq:yt-leading} with $\epsilon_j$ given by Eq.~\eqref{eq:beta-model} yields
\eq{
	\yt^{-1}=[(\omega\beta+V)^2-\lambda^2]/\lambda^2\,,
	\label{eq:yt-beta}
}
where $\lambda=1+\sqrt{1-\beta^2}$.  Setting $\yt^{-1}=0$ in Eq.~\eqref{eq:yt-beta}  yields correctly a single~\cite{ganeshan2015nearest} ME trajectory, given by 
\eq{
	\omega_\mathrm{ME}^{\pd} = \beta^{-1}[1+\sqrt{1-\beta^2}-V]\overset{\beta\ll 1}{\sim}\beta^{-1}(2-V)\,.
	\label{eq:ME-beta}
}
Eq.~\eqref{eq:ME-beta} is in qualitative agreement with the exactly known MEs~\cite{ganeshan2015nearest}, and asymptotically exact for $\beta\ll 1$; recovering as $\beta \to 0$ the  AAH model result~\cite{aubry1980analyticity} that all states undergo a one-shot transition at the critical $V=2$. Additionally, from the average DoS we obtain the localised band-edges as $\bel{\pm}= \pm V/(1\mp \beta)$, and hence via Eq.\ \eqref{eq:ME-beta} that $V_\pm=(1\pm\beta)\lambda$. Combining these results shows that localised states reside in ${\omega_\mathrm{ME}<\omega<\bel{+}}$ for $V\in(V_{-},V_{+})$,  and  $\bel{-}<\omega<\bel{+}$ for ${V>V_+}$  (where all states are localised). Analysing Eq.~\eqref{eq:deltat-leading} also shows that extended states exist in $\bed{-}<\omega<\omega_\mathrm{ME}$ for  $V\in(V_{-},V_{+})$, and $\bed{-}<\omega<\bed{+}$ for $V\in (0,V_-)$ (where all states are extended);  with  $\bed{\pm}$ the spectral edges obtained from the vanishing of $\deltat(\omega)$  via Eq.~\eqref{eq:deltat-leading}, given by $\pm\beta^2\bed{\pm} = \lambda\mp V\beta-\sqrt{(1-\beta^2)\lambda^2\mp 2V\beta\lambda}$ (and evolving smoothly into $\bel{\pm}$ at $V=V_{\pm}$).  Thus the collective information of  $\omega_\mathrm{ME}$, $\beld{\pm}$ and $V_\pm$ maps out the entire localisation phase diagram of the model in the $(V,\omega)$-plane. The  above results, as well the resultant phase diagram of the model, are summarised in Fig.~\ref{fig:Delta-y-typical-leading}(a).

Turning to the $l=2$ mosaic models Eq.~\eqref{eq:mosaic-model}, Eq.~\eqref{eq:yt-leading} gives
\eq{
	\yt^{-1}=(V\vert\omega\vert-2)/2\,,
	\label{eq:yt-mosaic}
}
with localised band-edges $\bel{\pm}=\pm V$. From $\yt^{-1}=0$ the model thus hosts two (symmetric in $\omega$) MEs given by 
\eq{\omega_{\mathrm{ME},\pm}^{\pd}=\pm 2/V\,, \label{eq:ME-mosaic}} which recovers precisely the exact result~\cite{wang2020onedimensional}. Note that the MEs never exit the spectrum, some states always being extended (such that $V_{+}$$\to$$ \infty$);
while the intersection of $\omega_{\mathrm{ME},\pm}$ and $\bel{\pm}$ gives $V_-=\sqrt{2}$. Localised states thus arise for $V>V_-$ in the regimes $\bel{-}<\omega<\omega_{\mathrm{ME},-}$ and $\omega_{\mathrm{ME},+}<\omega<\bel{+}$. For $V<V_-$, the spectrum  has solely extended states, with the band-edges obtained from the vanishing of $\deltat(\omega)$ (Eq.~\eqref{eq:deltat-leading}) as $\bed{\pm}=\pm2/\sqrt{4-V^2}$. The model thus hosts extended states in the regime $\bed{-}<\omega<\bed{+}$ for $V<V_-$ and $\omega_{\mathrm{ME},-}<\omega<\omega_{\mathrm{ME},+}$ for $V>V_-$. Similarly to the $\beta$-model, these results, together with the resultant phase diagram in the $(V,\omega)$-plane, are summarised  in Fig.~\ref{fig:Delta-y-typical-leading}(b).

\begin{figure}
\includegraphics[width=\linewidth]{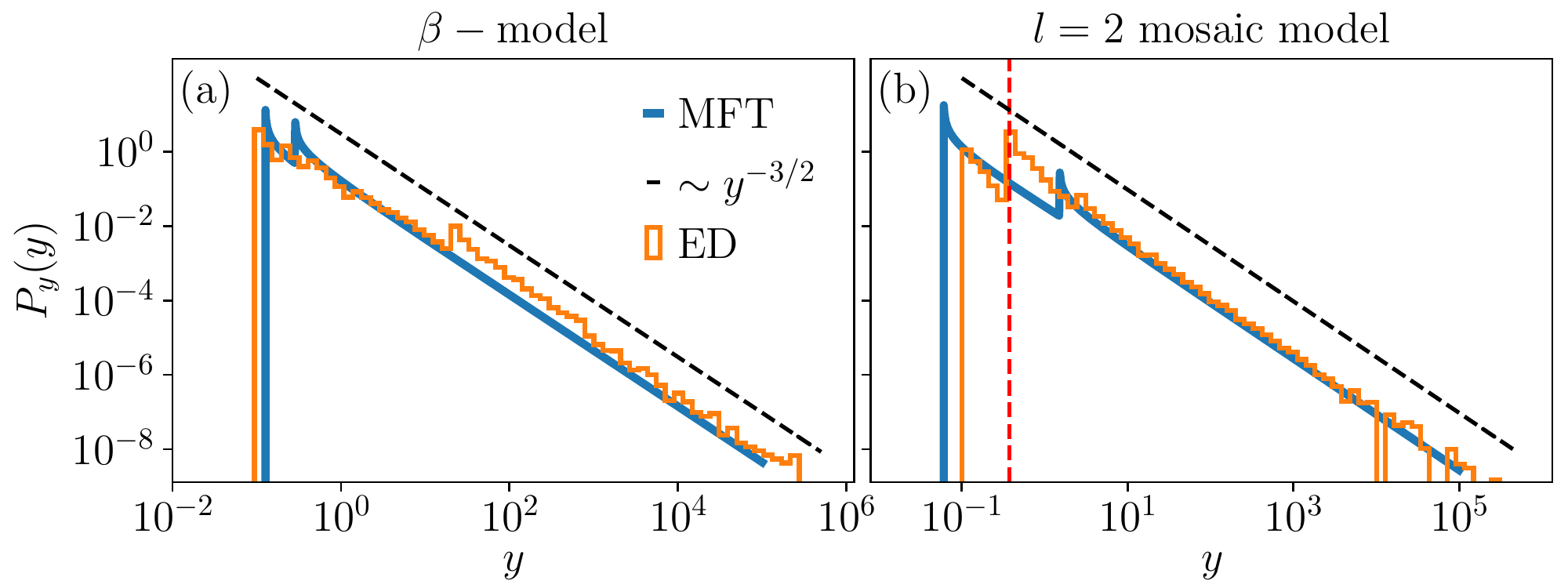}
\caption{Probability distributions of $y$ for the end site in the localised phase, for (a) the $\beta$-model with $\beta=0.2$, $V=3$, $\omega=0$; and (b) the $l=2$ mosaic model with $V=3$ and $\omega=2$. Blue lines show results from the leading-order theory. They are in excellent agreement with numerical data (shown in orange) obtained from exact diagonalisation (ED) for $L=2500$ sites with statistics obtained over 5000 $\phi$-values.  The red dashed vertical line in (b) shows the $\delta$-function contribution,  arising from even end sites~\cite{supp}. The black dashed lines in both panels show a $y^{-3/2}$ tail, confirming that $P_y(y)$ has a characteristic L\'evy tail.}
\label{fig:loc_dists_vs_numerics}
\end{figure}

 For each class of models, Eqs.\ \eqref{eq:yt-beta},\eqref{eq:yt-mosaic} show that $\yt$ diverges as $\yt \sim (\omega -\omega_{\mathrm{ME}})^{-1}$ on approaching the  ME from the localised side (with $\yt \sim (V-2)^{-1}$ as $V\to 2^{+}$ in the $\omega$-independent AAH limit). As this divergence is proportional to that of the localisation length $\xi(\omega)$~\cite{roy2020localisation}, $\xi(\omega)$ thus diverges with a critical exponent of $\nu =1$; which likewise agrees with the exactly known Lyapunov exponents for the mosaic~\cite{wang2020onedimensional} and AAH~\cite{thouless1983bandwidths} models.

Finally, while our natural focus has been on MEs, the theory also enables the self-consistent distributions of
$y_{0}$ and $\Delta_{0}$ to be obtained. Here we simply make some brief remarks about the distribution $P_{y}(y)$ of $y\equiv y_0$; which within the leading-order theory is given from Eq.~\eqref{eq:y0-leading} by 
\eq{
	P_y(y) = \Big\langle\Big\langle\delta\Big(y-\frac{J^2(1+\yt)}{(\omega-V\epsilon_1(\phi))^2}\Big)\Big\rangle\Big\rangle\,.
	\label{eq:Pydistloc}
}
The analytic results for $P_y(y)$~\cite{supp} are rather unwieldy, so we show them graphically in Fig.~\ref{fig:loc_dists_vs_numerics} for representative $(V,\omega)$ points for the two models; and compare them to results obtained from exact diagonalisation, with which excellent agreement is seen. Two  notable points to take away from the analytic expressions are however  that (i) the distributions have a $\propto y^{-3/2}$ power-law (L\'evy) tail; and (ii) the support of the distributions have a sharp lower cutoff, which arises because the $\epsilon_j$'s have a bounded distribution.  We add that the L\'evy tail in $P_{y}(y)$ seems quite universal, as it arises also in Anderson localisation in disordered systems in the presence of both uncorrelated~\cite{abou-chacra1973self} as well as maximally correlated disorder~\cite{roy2020localisation}.

In summary, we have introduced a self-consistent theory for MEs in quasiperiodic chains with nearest-neighbour hoppings. The theoretical framework is model-independent, and its efficacy was demonstrated using two different classes of quasiperiodic models. The central object of interest is the imaginary part of the local self-energy, which acts as a probabilistic order parameter for a localisation transition.  MEs arising from the theory, and the localisation phase diagram in the space of Hamiltonian parameters and energy, were found to be in very good agreement with previous numerical and analytical results on the same families of models.

The present work suggests natural directions for further research. Here, the continued fraction Eq.~\eqref{eq:S0-contfrac} for $S_0(\omega)$ has been analysed self-consistently upon truncating it at leading order. Generalising the theory to arbitrarily high orders, analysing the continued fraction's convergence, and connecting these approaches both conceptually and mathematically, forms the subject of a forthcoming work~\cite{duthie2020localisation}. Using the $\omega$-dependent propagators extracted here, to obtain analytical insights into some of the non-equilibrium  dynamics of such systems in the time-domain~\cite{roy2020imbalance}, is another interesting avenue for the future. Finally, developing a self-consistent theory of many-body localisation in quasiperiodic systems on the Fock space, wherein the quasiperiodicity of the potentials will generate strong correlations in the Fock-space disorder~\cite{logan2019many,roy2020fock}, remains a challenge.

\begin{acknowledgments}
We thank I. Creed  for helpful discussions. We also thank the EPSRC for support, under Grant No. EP/L015722/1 for the TMCS Centre for Doctoral Training, and Grant No. EP/S020527/1.
\end{acknowledgments}

\bibliography{refs}


\clearpage
{
\onecolumngrid
\begin{center}
\textbf{Supplementary material: Self-consistent theory of mobility edges in quasiperiodic chains}\\
Alexander Duthie, Sthitadhi Roy and David E. Logan
\end{center}
\bigskip
}

\twocolumngrid

\setcounter{equation}{0}
\renewcommand{\theequation}{S\arabic{equation}}
\setcounter{figure}{0}
\renewcommand{\thefigure}{S\arabic{figure}}
\setcounter{page}{1}
\renewcommand{\thepage}{S\arabic{page}}


\subsection{Density of localised states}

Let $i_{0}$ denote a generic end site of the semi-infinite chain considered  (denoted simply by $0$ in the main text).
Recall first that $\Delta_{i_{0}}(\w)=\pi J^{2}D_{i_{0}+1}(\w)$, with $D_{i_{0}+1}(\w)$  $=-\mathrm{Im}G_{i_{0}+1}^{(i_{0})}(\w)$ the LDoS for site $i_{0}+1$. Our self-consistent theory for the extended phase yields $\deltat(\w) \geq 0$, with $\deltat(\w)=\pi J^{2}D_{\mathrm{typ}}(\w)$ thus a direct measure of the typical LDoS. $\deltat(\w)$ can vanish for two reasons. The first, physically non-trivial reason, is when $\w$ approaches a ME;  the second corresponds simply to $\w$ approaching a spectral band edge. Each of these is contained in the solutions to Eq.\ \eqref{eq:deltat-leading} (with resultant MEs, and band edges $\bed{\pm}$, given in the main text).

In the localised phase by contrast, the central quantity is $\yt(\w)$, which diverges as $\w$ approaches a ME. The band edges in this regime (denoted by $\bel{\pm}$) can be  inferred  from the averaged DoS, given by
\begin{equation}
\label{eq:A1}
D(\w)~=~\big\langle\big\langle
D_{i_{0}}(\w)
\big\rangle\big\rangle
~=~\frac{1}{N}\sum_{i_{0}}\int_{0}^{2\pi}\frac{d\phi}{2\pi}~D_{i_{0}}(\w;\phi)
\end{equation}
with the average over both $\phi \in [0,2\pi]$ and end sites $i_{0}$ (the number of which is denoted by $N$). For the leading-order theory considered here, the local propagator  $G_{i_{0}}^{(i_{0}-1)}(\w) =[\w+i\eta-V\epsilon_{i_{0}}+i\deltat(\w)]^{-1}$.
Hence, for the regime of localised states in which $\deltat(\w) \propto \eta =0^{+}$,  we have  $D(\w)\equiv D_{L}(\w) =\langle\langle\delta(\w -V\epsilon_{i_{0}})\rangle\rangle$, i.e.\
\begin{equation}
\label{eq:A2}
D_{L}^{\pd}(\w) ~=~ \frac{1}{N}\sum_{i_{0}}^{}\int_{0}^{2\pi}\frac{d\phi}{2\pi}~\delta\big(\w -V\epsilon_{i_{0}}^{\pd}(\phi)\big).
\end{equation}

For the $\beta$-model~\cite{ganeshan2015nearest} (with $\epsilon_{i_{0}}$ from Eq.\ \eqref {eq:beta-model}), the $\phi$-integral in Eq.\ \eqref{eq:A2} is independent of the site index $i_{0}$ (whence the site average $N^{-1}\sum_{i_{0}}$ is in effect redundant); and evaluation of Eq.\ \eqref{eq:A2} gives
\begin{equation}
\label{eq:A3}
D_{L}^{\pd}(\w) ~=~\frac{1}{\pi\left(1+\frac{\beta \w}{V}\right)\sqrt{
\big(\beta \w+V \big)^{2} -\w^{2}}},
\end{equation}
holding for $\w^{2} \leq (\beta\w+V)^{2}$, with the equality giving the band edges $\bel{\pm}=\pm V/(1\mp \beta)$. Note that for $V=V_{-}$ [$V_{+}$], below [above] which $V$ all states are extended [localised],  and where the ME coincides with a band edge, the band edge  $\bel{+}$ [$\bel{-}$] correctly coincides with  $\bed{+}$ [$\bed{-}$] arising from the vanishing of Eq.\ \eqref{eq:deltat-leading} for $\deltat(\w)$. For the $l=2$ mosaic model~\cite{wang2020onedimensional}  (Eq.\ \ref {eq:mosaic-model}), half the end sites $i_{0}$ have an $\epsilon_{i_{0}}$ of AAH form while the remainder have $\epsilon_{i_{0}}=0$,  with $\phi$-integrals in each case again independent of $i_{0}$; yielding
\begin{equation}
\label{eq:A5}
D_{L}^{\pd}(\w) ~=~\frac{1}{2\pi\sqrt{V^{2} -\w^{2}}}~+~\tfrac{1}{2}\delta (\w)
\end{equation}
with $\bel{\pm}=\pm V$. Once again, for $V=V_{-}$  the MEs and band edges again coincide, with 
$\omega_{\mathrm{ME},\pm} =\bel{\pm}=\bed{\pm}$ (recall that $V_{+}\to \infty$ for this model, as some states are
always extended).

\subsection{Analytic results for $P_{y}(y)$}

In the main text, we demonstrated graphically that the distributions $P_y(y)$ have a signature $\propto y^{-3/2}$ L\'evy tail (see Fig.~3). Here we give the analytical derivations and resulting expressions for $P_y(y)$.

The starting point is Eq.\ \eqref{eq:Pydistloc}, which can be re-expressed as 
\begin{equation}
P_y(y) = \frac{1}{2\pi N}\sum_{i_0}\sum_{\phi^\ast}\vert\partial_\phi^{\pd} f\vert^{-1}\big\vert_{\phi=\phi^\ast},
\label{eq:py-sm}
\end{equation}
where 
\eq{
f = \frac{J^2(1+y_\mathrm{typ})}{(\omega-V\epsilon_1^{\pd}(\phi))^2}
\label{eq:f}
} 
and the set of $\phi^\ast$ values are defined as the solutions to $f(\phi^\ast)=y$. In the following we set $J=1$ for brevity.  Note that the functional form of $f$ in Eq.~\eqref{eq:f} in general generates two solutions, $\phi^\ast_\pm$, given by 
\eq{
    V\epsilon_1^{\pd}(\phi_\pm^\ast)=\omega\mp  \sqrt{\frac{1+y_\mathrm{typ}}{y}},
\label{eq:phistar}
}
such that $P_{y}(y)$ can be written as a sum of two contributions,
\eq{
P_y^{\pd}(y) &= P_y^+(y) + P_y^-(y)
\label{eq:Pplusminus}
}
with $P_y^\pm(y)=\frac{1}{2\pi N}\sum_{i_0}\vert\partial_\phi f\vert^{-1}\vert_{\phi=\phi^\ast_\pm}$.  It is also useful to notice that since the magnitude of $\epsilon_1(\phi)$ is bounded from above, the support of the distribution $P_y(y)$ is bounded from below. $P_y(y)$ is thus cut off at small values of $y$, as can be seen in Fig.~3.

We start with the $\beta$-model, where we give the results explicitly only for the band-centre, $\omega=0$. Using Eqs.~\eqref{eq:py-sm}-\eqref{eq:phistar} with $\epsilon_i(\phi)$ given by Eq.\ \eqref{eq:beta-model}, we obtain
\begin{align}
\begin{split}
    P_y^\pm(y) = &\frac{y^{-3/2}}{2\pi}\frac{V\sqrt{1+\yt}}{V\pm\beta\sqrt{\frac{1+\yt}{y}}} \\ &\times\left\{ \left(V\pm\beta\sqrt{\frac{1+\yt}{y}}\right)^2 - \frac{1+\yt}{y}   \right\}^{-1/2},
\end{split}
\label{eq:beta-dist}
\end{align}
with the support of $P_y^\pm(y)$ residing in
\eq{
 y\geq \frac{(1+\yt)}{V^2}(1\mp\beta)^2. 
}
Most importantly, for $y \gg \yt$, Eq.~\eqref{eq:beta-dist} gives
\eq{
P_y^\pm(y)\overset{y\gg \yt}{\sim}\frac{\sqrt{1+\yt}}{2\pi V}y^{-3/2},
}
which clearly shows the characteristic L\'evy tail.

Turning to the $l=2$ mosaic model, we note that there can be two kinds of end-sites, with odd and even $i_0$ respectively. Hence, each of the terms $P_y^\pm$ can be expressed as a sum of two terms
\eq{
P_y^\pm(y) = P_y^{\mathrm{even},\pm}(y) + P_y^{\mathrm{odd},\pm}(y).
}
For even $i_0$, $\epsilon_1 = 0$, and
\eq{
P_y^{\mathrm{even},\pm}(y) = \frac{1}{4}\delta\left(y-\frac{1+\yt}{\omega^2}\right).
\label{eq:mosaic-py-even}
}
This is the $\delta$-function contribution shown by the red dashed vertical line in Fig.~3(b).
Using the potential in Eq.\ \eqref{eq:mosaic-model} for odd $i_0$ in Eqs.~\eqref{eq:py-sm}-\eqref{eq:phistar} we obtain
\eq{
    P_y^{\mathrm{odd},\pm}(y) = \frac{1}{4\pi}y^{-3/2}\sqrt{\frac{1+y_t}{V^2-R_\pm^2}}~;~~R_\pm=\omega\pm\sqrt{\frac{1+y_t}{y}},
    \label{eq:mosaic-py-odd}
}
with the distributions supported on $y\geq (1 +\yt)/(\omega\mp V)^2$.
The sum of the even and odd contributions, Eqs.~\eqref{eq:mosaic-py-even} and \eqref{eq:mosaic-py-odd}, comprise the full distribution. Again, for $y\gg \yt$, it is readily seen from Eqs.~\eqref{eq:mosaic-py-even} and \eqref{eq:mosaic-py-odd} that $P_y(y)$ has a form
\eq{
P_y^{\pd}(y) \overset{y\gg \yt}{\sim} \frac{1}{2\pi}\sqrt{\frac{1+\yt}{V^2-\omega^2}}~~y^{-3/2},
}
likewise displaying the characteristic L\'evy tail.

\end{document}